\documentclass[vecphys]{svmult}

\usepackage{makeidx}         
\usepackage{graphicx}        
\usepackage{multicol}        
\usepackage[bottom]{footmisc}

\makeindex             

\begin{document}

\title*{Global dynamics of coupled standard maps}

\author{T.~Manos\inst{1,2}, Ch.~Skokos\inst{3} \and
T.~Bountis\inst{1}} \institute{Center for Research and Applications
of Nonlinear Systems (CRANS), Department of Mathematics, University
of Patras, GR--26500, Greece.\\
\texttt{thanosm@master.math.upatras.gr},
\texttt{bountis@math.upatras.gr} \and
Observatoire Astronomique de
Marseille--Provence (OAMP), 2 Place Le Verrier, F--13248
Marseille, C\'edex 04, France.\\ \and Astronomie et Syst\`{e}mes
Dynamiques, IMCCE, Observatoire de Paris,
\\ 77 Av. Denfert--Rochereau, F--75014, Paris, France. \texttt{hskokos@imcce.fr} }

\maketitle
\begin{abstract}
Understanding the dynamics of multi--dimensional conservative
dynamical systems (Hamiltonian flows or symplectic maps) is a
fundamental issue of nonlinear science. The Generalized ALignment
Index (GALI), which was recently introduced and applied successfully
for the distinction between regular and chaotic motion in
Hamiltonian systems \cite{sk:6}, is an ideal tool for this purpose.
In the present paper we make a first step towards the dynamical
study of multi--dimensional maps, by obtaining some interesting
results for a 4--dimensional (4D) symplectic map consisting of $N=2$
coupled standard maps \cite{Kan:1}. In particular, using the new
GALI$_3$ and GALI$_4$ indices, we compute the percentages of regular
and chaotic motion of the map equally reliably but much faster than
previously used indices, like GALI$_2$ (known in the literature as
SALI).
\end{abstract}

\vspace{-1.25cm}
\section{Definition and behavior of GALI}
\label{GALI_def}
\vspace{-0.25 cm}
Let us first briefly recall the definition of GALI and its behavior
for regular and chaotic motion, adjusting the results obtained in
\cite{sk:6} to symplectic maps. Considering of a $2N$--dimensional map,
we follow the evolution of an orbit (using the equations of the map)
together with $k$ initially linearly independent deviation vectors of
this orbit $\overrightarrow{\nu}_{1},\overrightarrow{\nu}_{2},...,
\overrightarrow{\nu}_{k}$ with $2\leq k \leq 2N$ (using the equations
of the tangent map). The Generalized Alignment Index of order $k$ is
defined as the norm of the wedge or exterior product of the $k$ unit
deviation vectors:
\begin{equation}\label{GALI:0}
    GALI_{k}(i)=\parallel \hat{\nu}_{1}(i)\wedge \hat{\nu}_{2}(i)
    \wedge ... \wedge \hat{\nu}_{k}(i) \parallel
\end{equation}
and corresponds to the volume of the generalized parallelepiped, whose
edges are these $k$ vectors. We note that the hat ($\,\hat{}\,$) over
a vector denotes that it is of unit magnitude and that $i$ is the
discrete time.

In the case of a chaotic orbit all deviation vectors tend to become
linearly dependent, aligning in the direction of the eigenvector which
corresponds to the maximal Lyapunov exponent and GALI$_{k}$ tends to
zero following an exponential law $ \sim
e^{-[(\sigma_{1}-\sigma_{2})+(\sigma_{1}-\sigma_{3})
+...+(\sigma_{1}-\sigma_{k})]i}$, where $\sigma_1, \ldots, \sigma_k$
are approximations of the first $k$ largest Lyapunov exponents. In the
case of regular motion on the other hand, all deviation vectors tend
to fall on the $N$--dimensional tangent space of the torus on which
the motion lies. Thus, if we start with $k\leq N$ general deviation
vectors they will remain linearly independent on the $N$--dimensional
tangent space of the torus, since there is no particular reason for
them to become aligned. As a consequence GALI$_{k}$ remains
practically constant for $k\leq N$. On the other hand, GALI$_{k}$
tends to zero for $k>N$, since some deviation vectors will eventually
become linearly dependent, following a particular power law, i.~e.~$GALI_{k}(i) \sim i^{2(N-k)}$.

\vspace{-0.5 cm}
\section{Dynamical study of a 4D standard map}

As a model for our study we consider the 4D symplectic map:
\begin{equation}\label{4Dmap}
\begin{array}{lll}
  x_{1}' &=& x_{1} + x_{2}' \qquad x_{2}' = x_{2} +
  \frac{K}{2\pi}\sin(2\pi x_{1}) -
  \frac{B}{2\pi}\sin[2\pi(x_{3}-x_{1})]\\
  x_{3}' &=& x_{3} + x_{4}' \qquad x_{4}' = x_{4} +
  \frac{K}{2\pi}\sin(2\pi x_{3}) -
  \frac{B}{2\pi}\sin[2\pi(x_{1}-x_{3})]\\
  \end{array} \, (mod\,1),
\end{equation}
which consists of two coupled standard maps \cite{Kan:1} and is a
typical nonlinear system, in which regions of chaotic and
regular dynamics are found to coexist. In our study we fix the
parameters of the map (\ref{4Dmap}) to $K=0.5$ and $B=0.05$.
\begin{figure}[t]
\centering
\includegraphics[height=5.9cm]{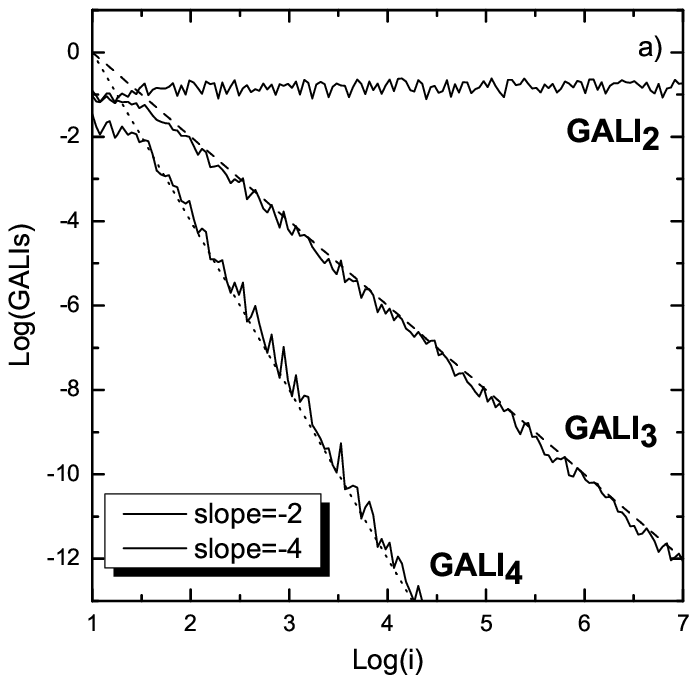}\hspace{-0.5 cm}
\includegraphics[height=5.9cm]{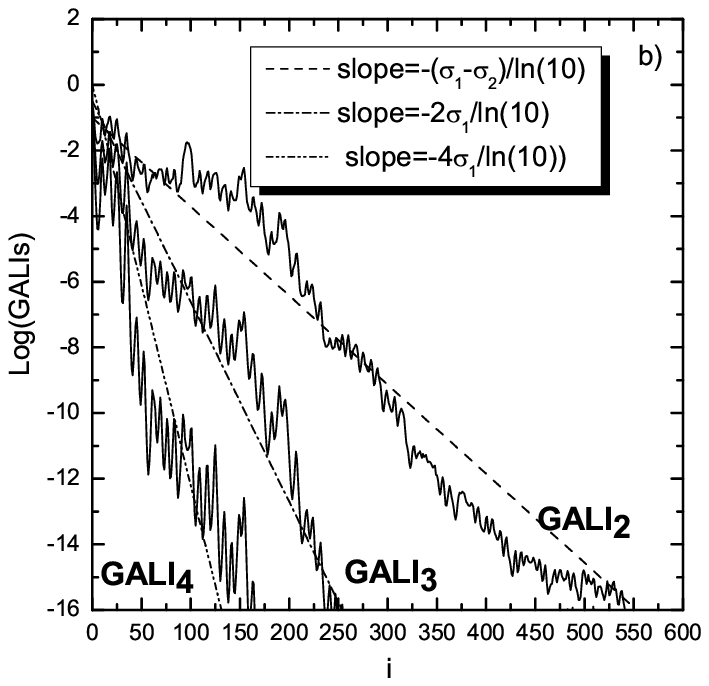}\vspace{-0.5 cm}
\caption{The evolution of GALI$_k$, $k=2,3,4$, with respect to the
number of iterations $i$ for a) the regular orbit R and b) the chaotic
orbit C. The plotted lines correspond to functions proportional to
$n^{-2}$, $n^{-4}$ in a) and to $ e^{-(\sigma_1 - \sigma_2)i}$, $e^{-2
\sigma_1 i}$, $e^{-4\sigma_1 i} $ for $\sigma_{1}= 0.070$,
$\sigma_{2}= 0.008$ in b).} \label{4D_GALI_beh} \vspace{-0.5 cm}
\end{figure}

In Fig.~\ref{4D_GALI_beh}, we show the behavior of GALIs for two
different orbits: a regular orbit R with initial conditions
$(x_1,x_2,x_3,x_4)=(0.55,0.10,0.54,0.01)$ (Fig.~\ref{4D_GALI_beh}a),
and a chaotic orbit C with initial conditions
$(x_1,x_2,x_3,x_4)=(0.55,0.10,0.005,0.01)$
(Fig.~\ref{4D_GALI_beh}b). The positive Lyapunov exponents of orbit C
were found to be $\sigma_{1}\approx 0.070$, $\sigma_{2}\approx
0.008$. From the results of Fig.~\ref{4D_GALI_beh} we see that the
evolution of GALIs is described very well by the theoretically
obtained approximations presented in Sect. \ref{GALI_def}.
\begin{figure}[t]
\centering
\includegraphics[width=5.9cm]{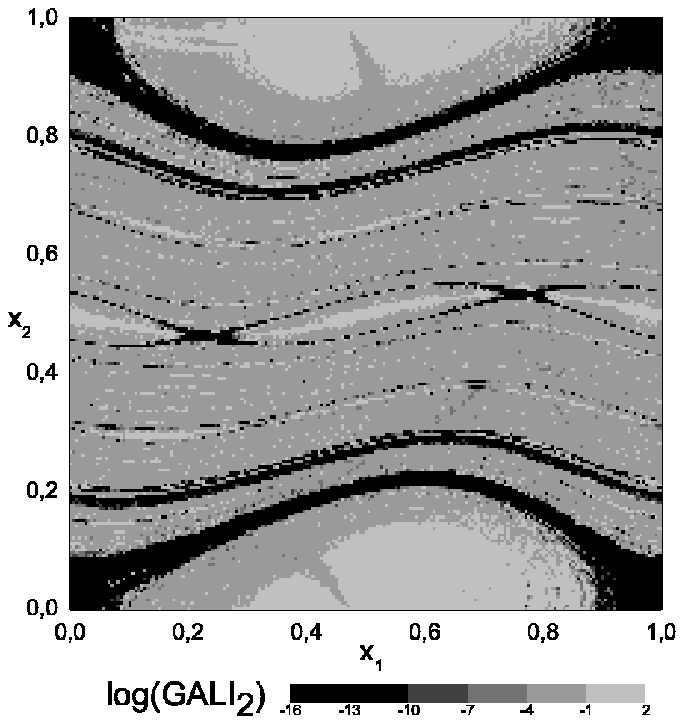}\hspace{-0.5cm}
\includegraphics[width=6.1cm]{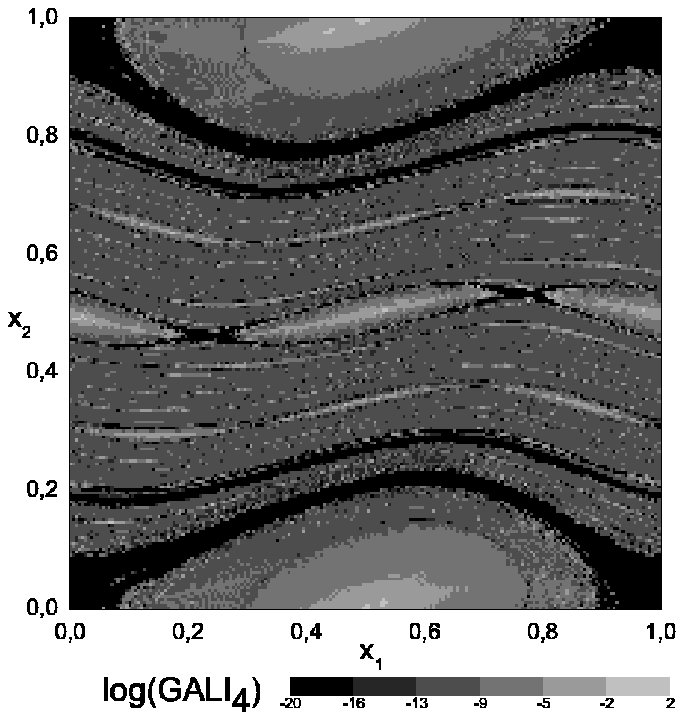}\vspace{-0.4 cm}
\caption{Regions of different values of the GALI$_2$ (left panel)
and GALI$_4$ (right panel) for a grid of $500\times 500$
initial conditions on the subspace $x_3= 0.54$, $x_4=0.01$ of map (\ref{4Dmap})
for $K=0.5$ and $B=0.05$.} \label{scan:1} \vspace{-0.5 cm}
\end{figure}

Let us now turn our attention to the study of the global dynamics of
map (\ref{4Dmap}). From the results
Fig.~\ref{4D_GALI_beh} we conclude that in the case of 4D maps,
GALI$_2$ has different behavior for regular and chaotic orbits. In
particular, GALI$_2$ tends exponentially to zero for chaotic orbits
(GALI$_2 \sim e^{-(\sigma_1 - \sigma_2)i}$) while it fluctuates around
non--zero values for regular orbits. This difference in the behavior
of the index can be used to obtain a clear distinction between regular
and chaotic orbits. Let us illustrate this by following up to $i=4000$
iterations, all orbits whose initial conditions lie on a
2--dimensional grid of $500\times 500$ equally spaced points on the
subspace $x_3= 0.54$, $x_4=0.01$, of the 4--dimensional
phase space of the map (\ref{4Dmap}), attributing to each grid point a
color according to the value of GALI$_2$ at the end of the
evolution. If GALI$_2$ of an orbit becomes less than $10^{-10}$ for
$i<4000$ the evolution of the orbit is stopped, its GALI$_2$ value is
registered and the orbit is characterized as chaotic. The outcome of
this experiment is presented in the left panel of Fig.~\ref{scan:1}.
\begin{figure}[t]
\centering
\includegraphics[width=5.9cm]{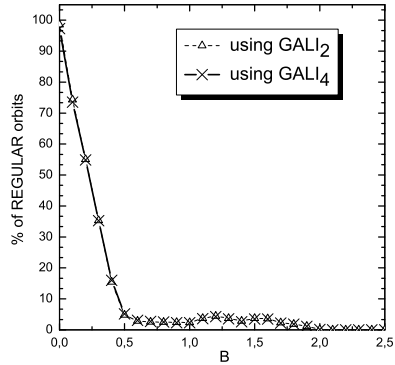}\vspace{-0.5 cm}
\caption{Percentages of regular orbits on the subspace $x_3= 0.54$,
$x_4=0.01$ of map (\ref{4Dmap}) for $K=0.5$, as a function of the
parameter $B\in [0,2.5]$.} \label{perc}\vspace{-0.5 cm}
\end{figure}

But also GALI$_4$ can be used for discriminating regular and chaotic
motion.  From the theoretical predictions for the evolution of
GALI$_4$, we see that after $i=1000$ iterations the value of GALI$_4$
of a regular orbit should become of the order of $10^{-16}$, since
GALI$_4 \sim t^{-4}$, although the results of Fig.~\ref{4D_GALI_beh}
show that more iterations are needed for this threshold to be reached,
due to an initial transient time where GALI$_4$ does not decrease
significantly. On the other hand, for a chaotic orbit GALI$_4$ has
already reached extremely small values at $i=1000$ due to its
exponential decay (GALI$_4 \sim e^{-4\sigma_1 i}$). Thus, the global
dynamics of the system can be revealed as follows: we follow the
evolution of the same orbits as in the case of GALI$_2$ and register
for each orbit the value of GALI$_4$ after $i=1000$ iterations. All
orbits having values of GALI$_4$ significantly smaller than $10^{-16}$
 are characterized as chaotic, while
all others are considered as non--chaotic. In the right panel of
Fig.~\ref{scan:1} we present the outcome of this procedure.

From the results of Fig.~\ref{scan:1}, we see that both procedures,
using GALI$_2$ or GALI$_4$ as a chaos indicator, give the same
result for the global dynamics of the system, since in both cases
16\% of the orbits are characterized as chaotic. These orbits
correspond to the black colored areas n both panels of
Fig.~\ref{scan:1}. One important difference between the two
procedures is their computational efficiency. Even though GALI$_4$
requires the computation of four deviation vectors, instead of only
two that are needed for the evaluation of GALI$_2$, using GALI$_4$
we were able to get a clear dynamical `chart', not only for less
iterations of the map (1000 instead of 4000 needed for GALI$_2$), but
also in less CPU time. In particular, for the computation of the
data of the left panel of Fig.~\ref{scan:1} (using GALI$_2$) we
needed 1 hour of CPU time on an Athlon 64bit, 3.2GHz PC, while for
the data of the left panel of the same figure (using GALI$_4$) only
14 minutes of CPU time were needed.

Using the above--described method, both for GALI$_2$ and GALI$_4$,
we were able to compute very fast and accurately the percentages of
regular motion for several values of parameter $B$.  In
Fig.~\ref{perc} we plot the percentage of regular orbits for $B\in
[0,2.5]$ where $B$ varies with a step $\delta B = 0.1$. We see that
the two curves practically coincide, but using GALI$_2$ we needed
almost four times more CPU time. So, it becomes evident that a
well--tailored application of GALI$_k$, with $2<k$, can
significantly diminish the CPU time required for the detailed
`charting' of phase space regions, compared with that for GALI$_2$.

\vspace{-0.5 cm}
\section*{Acknowledgments}
\vspace{-0.5 cm}
T.~Manos was supported by the ``Karatheodory" graduate student
fellowship No B395 of the Univ. of Patras, the program ``Pythagoras
II" and the Marie Curie fellowship No HPMT-CT-2001-00338. Ch.~Skokos
was supported by the Marie Curie Intra--European Fellowship No
MEIF--CT--2006--025678.


\vspace{-0.5 cm}
\begin{thebibliography}{9}
\vspace{-0.5 cm}
\bibitem{sk:6} Ch. Skokos, T. Bountis and Ch. Antonopoulos,
\emph{Physica D}, \textbf{231}, 30, (2007).

\bibitem{Kan:1} H. Kantz and P. Grassberger, \emph{J. Phys. A:
Math. Gen}, \textbf{21} L127, (1988).

\end{thebibliography}
\end{document}